\documentclass[preprint,number,12pt]{elsarticle}
\usepackage{lineno}
\usepackage{amssymb}
\usepackage{soul}
\usepackage{xcolor}
\usepackage[utf8]{inputenc}
\usepackage{graphicx}
\usepackage{amsmath}
\usepackage{gensymb}

\journal{Nucl. Instr. Meth. Phys. Res. A}

\biboptions{sort&compress}

\makeatletter
\def\ps@pprintTitle{%
 \def\@oddhead{\footnotesize\itshape
 $\copyright$ 2023.  This manuscript version is made available under the CC-BY-NC-ND 4.0 license. }
 \let\@evenhead\@empty
 \def\@oddfoot{\footnotesize\itshape
 Nucl. Instr. Meth. Phys. Res. A: Submitted Nov. 17, 2022; Accepted Jan. 22, 2023}%
 \let\@evenfoot\@oddfoot}
\makeatother

\begin{document}

\begin{frontmatter}
\title{Evaluation of Light Collection from Highly Scattering Media using Wavelength-Shifting Fibers}

\author[um]{Andrew S. Wilhelm\corref{cor1}}
  \ead{andhelm@umich.edu}

\author[psu]{Garrett Wendel}
    \ead{gmw5164@psu.edu}
    
\author[mtu]{Brandon Collins}
    \ead{bgcollin@mtu.edu}    
    
\author[psu]{Doug Cowen}
    \ead{dfc13@psu.edu}    

\author[um]{Igor Jovanovic}%
    \ead{ijov@umich.edu}

\cortext[cor1]{Corresponding Author}

\address[um]{%
 Department of Nuclear Engineering and Radiological Sciences, University of Michigan, Ann Arbor, MI 48105, USA\\
}%

\address[psu]{%
 Department of Physics, Pennsylvania State University, State College, PA 16801, USA\\
}%

\address[mtu]{%
 Department of Mechanical Engineering, Michigan Technological University, Houghton, MI 49931, USA\\
}%

\begin{abstract}
    Opaque scintillators are designed to have a short scattering length such that scintillation photons are probabilistically confined to a small region of space about their origin.  The benefit of this feature is that information on the interaction event topology can be recorded with greater fidelity than in traditional highly transparent media with sensors at large distances from the light production region.  Opaque scintillator detectors rely on wavelength-shifting fibers to extract the scintillation light; however, the efficiency of light collection has not yet been directly measured in experiment.  We measured the efficiency of light collection as a function of the optical parameters of an opaque liquid and the distance from the origin of the light to the fiber. We use the experimental data to validate a Monte Carlo model of light transport and collection and discuss a simple diffusion model that reproduces the results of Monte Carlo simulation with high fidelity. This combination of validated models has the potential for use in predictions of performance in various designs of future opaque scintillator detectors such as LiquidO.  
\end{abstract}


\end{frontmatter}


\section{Introduction}
\label{sec:intro}
All radiation detectors rely on the interaction of incident particles with matter, where the kinetic energy of primary or secondary particles is deposited in the instrumented detector volume.  In scintillation detectors, a fraction of the energy deposited by the interacting particle is emitted as light, which is then converted into an electrical signal via a photosensor such as a photomultiplier. The photosensor is usually placed outside the active volume, near its boundary.  In this case, the scintillation medium should be as transparent as possible so that the maximum fraction of generated light reaches the boundary and is detected by the photosensor.  A scintillation photon will often scatter or reflect multiple times at non-sensing detector boundaries before reaching the photosensor and, thus, the information available for position reconstruction in the detector active volume may be severely degraded or even entirely lost.  The locations of energy deposition within a detector volume arising from a single quantum of incident radiation are referred to collectively as the ``event topology.''   

The measured event topology in a radiation detector can be leveraged to infer the direction of origin of a radiation source~\cite{Dogan1990MultipleCamera} and to identify the type of interacting particle~\cite{IceCube}.  There are several ways to obtain event topology in a scintillation detector.  One method is to optically isolate segments of the detector, each of which then functions as a pixel~\cite{SANDD,Weinfurther2016LowCamera}; another is to use relative time-of-arrival of the optical photons at multiple photosensors, which is often employed in large detectors~\cite{Wonsak2018TopologicalDetectors} and in applications that require excellent timing resolution, such as Positron Emission Tomography (PET)~\cite{Kim2013TrendsTomography}. 

LiquidO~\cite{Cabrera2021} has been proposed as a method to localize scintillation photons by designing the detection medium to have a short scattering length, thus achieving virtual pixelization without the need to highly segment the detector volume.  A scintillation photon will travel only a short distance between scatters such that it is ``stochastically confined''~\cite{Cabrera2021, CabreraTalk2022}.  A challenge in using an opaque scintillator is that the light must be eventually extracted from the detection volume so that it can be sent to a photosensor and converted into an electrical signal. This can be achieved using a lattice of wavelength-shifting (WLS) fibers.  Scintillation photons are absorbed in the core of a WLS fiber and are isotropically reemitted with a longer wavelength, such that some of them meet the criterion for total internal reflection.  A photon captured in this manner can be transported through the fiber to the photosensor.  Event topology can then be reconstructed from the relative amount of light collected in multiple fibers that traverse the region near the origin of scintillation.  

An ideal opaque scintillator will have high light yield, a short scattering length, and a long absorption length.  Liquids with the latter two characteristics for photons in the visible range appear white and milky or waxy.  Currently, at least two materials are under consideration for LiquidO applications: (1)~NoWaSH, which is a mixture of linear alkyl benzene (LAB) and paraffin that reduces the scattering length~\cite{Buck2019} and (2) water-based liquid scintillator (WbLS), which can be tuned to have the desired optical properties~\cite{Yeh2011AApplications, Caravaca2020CharacterizationSeparation}.  Additionally, a scintillator with desired LiquidO properties may be created by adding nanoparticles (\textit{e.g.}, TiO$_2$) to a traditional liquid scintillator, though the stability of such a compound remains to be demonstrated.  

In order to design a LiquidO detector, an accurate model of light transport in the medium and collection into the fibers is required.   In this work, we experimentally measure the efficiency with which the light emitted by a highly scattering medium is collected by WLS fibers. We use this data to validate a Geant4~\cite{Agostinelli2003} model for the experiment, which offers the prospect for predicting the performance of future LiquidO detectors.     

\section{Experiment}
In order to measure light collection efficiency from an opaque liquid into WLS fibers as a function of distance between the light source and fiber, the precise origin of the light must be known.  To achieve this, we opted to use an inert (\textit{i.e.}, non-scintillating) liquid and directly inject the light at a known position using a fiber-coupled pulsed laser.  This allows the control of both the location of emission and intensity of the light in the opaque medium.  Additionally, since the time of arrival of laser pulses is known with great accuracy, the effect of spurious pulses arising from various background in this measurement can be greatly reduced. 

For this study, we chose to use dilutions of commercial cow milk as the opaque, non-scintillating liquid.  Milk has several salutary properties, one of which is that it has a longer absorption length than scattering length for the wavelengths of light under consideration \cite{Stocker2017BroadbandMilk}.  Another is that milk is water soluble, which means that it's optical properties can be easily tuned by mixing it with distilled water.  Additionally, milk is readily available and inexpensive.  Another candidate liquid was a TiO$_2$-water nanofluid.  However, in addition to agglomeration, TiO$_2$-water nanofluids suffer from sedimentation over time, at rates up to 5\% in an hour \cite{Brunelli2013AgglomerationWaters}.  Another attractive option is Intralipid, which has consistent optical properties batch-to-batch, stability on the timescale of years, and high scattering and low absorption \cite{Ninni2011Intralipid:Phantoms}.  However, it is much more expensive than milk.

\label{sec:exp}
\subsection{Apparatus}
\label{sec:app}

\begin{figure}
    \centering
    \includegraphics[width=\textwidth]{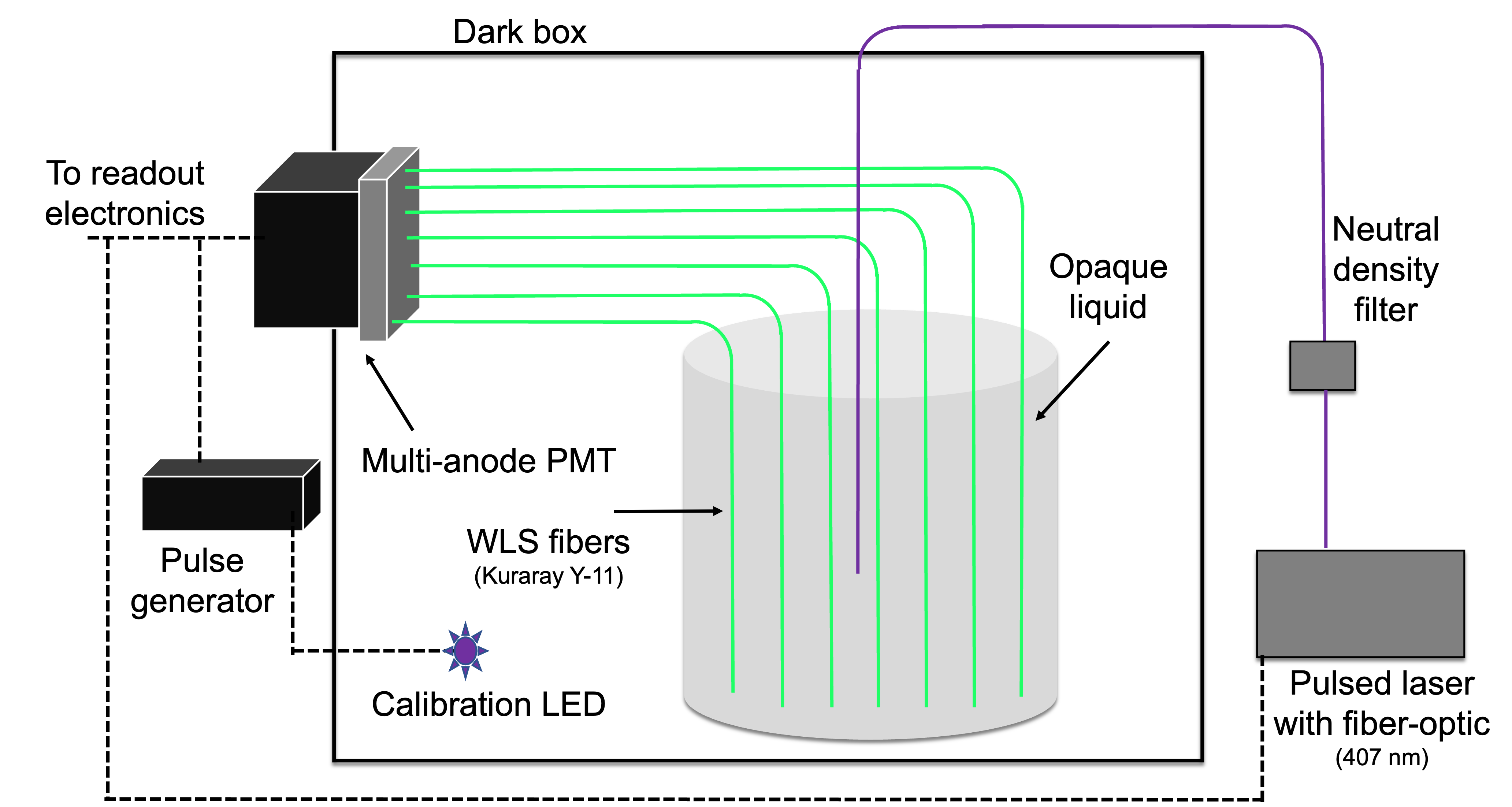}
    \caption{A sketch of the experimental apparatus.  The dashed interfaces are electrical, and the solid ones are optical.  The data acquisition system is triggered by the pulse generator during calibration and by the laser controller during experiments with laser pulses.}
    \label{fig:blockDiagram}
\end{figure}

\begin{figure}
    \centering
    \includegraphics[width=0.6\textwidth]{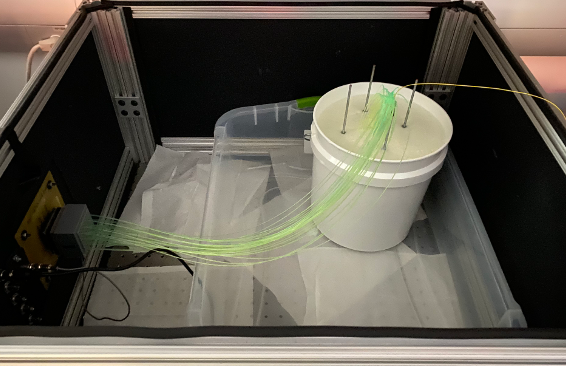}
    \caption{The experimental apparatus inside a dark box.  The yellow-jacketed fiber carries the light from the pulsed-laser to the center of the fiber array, which is submerged in 7500~ml of opaque liquid.  The fibers emerge from the liquid and terminate at individual pixels of the MAPMT.}
    \label{fig:darkBox}
\end{figure}

A schematic of the experimental setup is shown in Fig.~\ref{fig:blockDiagram}.  The experiment was conducted in a light-tight box (shown in Fig.~\ref{fig:darkBox}).  The container for the liquid was a 2-gallon, high-density polyethylene (HDPE) bucket.

The light source was a pulsed diode laser from NKT Photonics, operating at 407~nm.  The pulse duration is approximately 35~ps, and the output power is tunable.  The laser was coupled to a single-mode fiber after passing through neutral density filters to attenuate it.  In this experiment, we operated the laser at 10~kHz, such that each pulse contained approximately 18,000 photons after attenuation using the neutral density filters.  The laser power was measured prior to each experimental run using a Thorlabs PM101 power meter.  

The WLS fibers used were 1~m sections of Kuraray Y-11(200) (non-S type).  One end of each fiber terminated at the bottom of the vessel, with a gap of 10$\pm$3~mm to reduce reflections back into the fiber.  The other end was coupled to a pixel of a Hamamatsu H12700A multianode photomultiplier tube (MAPMT).  Each fiber end was cleaved using a Comptyco KPT-130 High Precision Cleaver
and then polished using a series of sandpapers with 1000--4000 grits.  Consistent contact between the fiber and photocathode was achieved using optical grease.  The H12700A has 64 pixels, but we used only 32 pixels in a checkerboard pattern to reduce cross-talk between neighboring channels.  The MAPMT was biased at $-$1100~V, and the waveforms were digitized with two CAEN V1730 boards.  Pulse integration was done onboard the V1730s using DPP-PSD firmware, with an integration window of 256~ns.  This was longer than necessary, as the decay time of the WLS fibers is 7.4~ns~\cite{Alekseev2022TheYS-2}.  The digitizer was triggered with a signal from the laser controller, regardless of whether a voltage threshold in that channel was met.  As a result, there were often no photoelectrons measured in a given integration window.  This approach enabled the determination of the average number of photoelectrons per pulse for each MAPMT pixel.        

\begin{figure}
    \centering
    \includegraphics[width=0.5\textwidth]{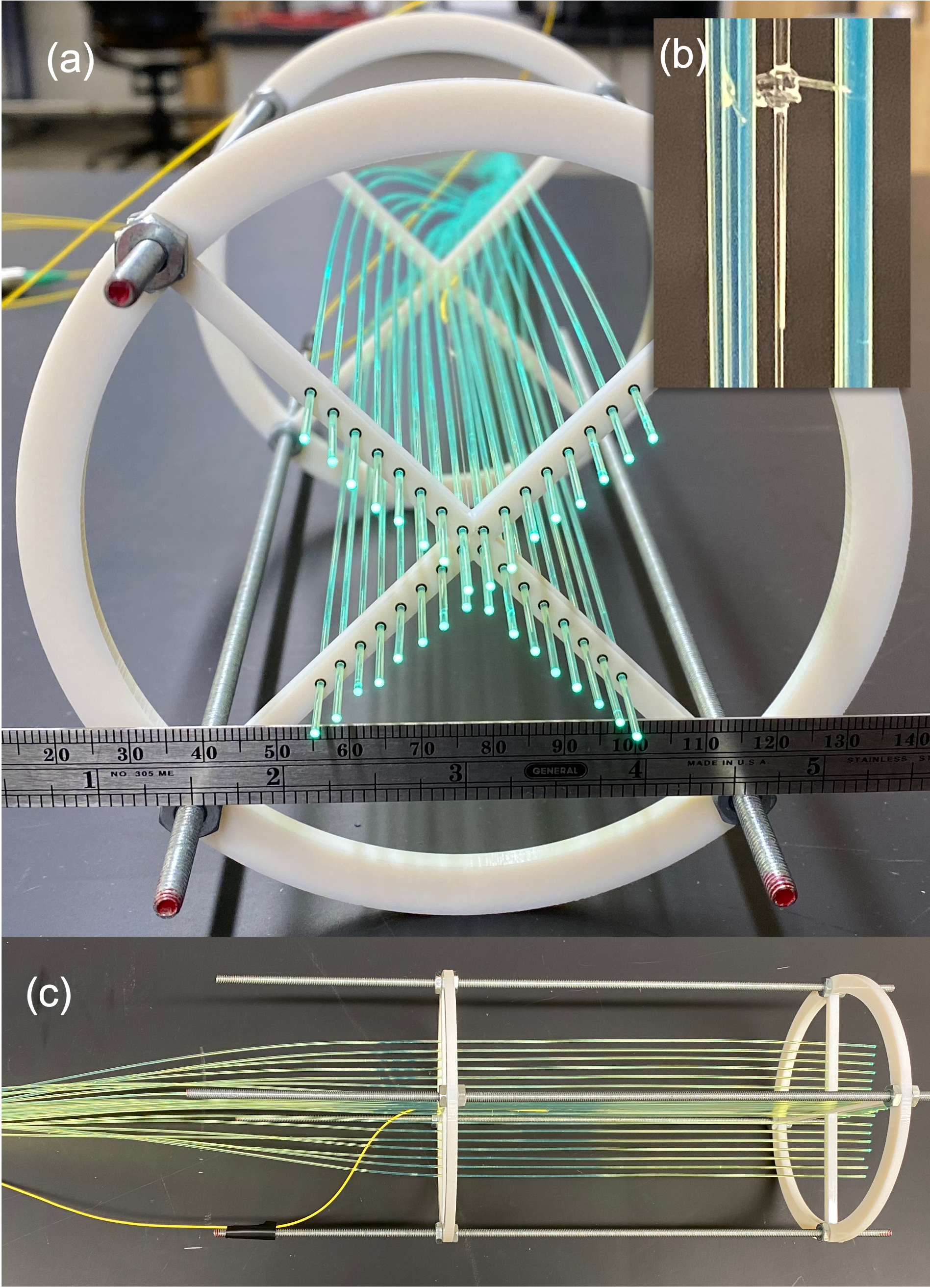}
    \caption{Views of the fiber holder with WLS fibers from (a)~bottom and (c)~side. The end of the laser-coupled single-mode fiber is shown in~(b).  A 0.2~mm diameter nylon monofilament is suspended vertically in the center of the array.  The single-mode fiber (ThorLabs SM300) is secured to the nylon monofilament by a small knot, made of more nylon monofilament, to maintain the central positioning of the light source.}
    \label{fig:fiberHolder}
\end{figure}

The fibers were held in place by a 3D-printed fixture; four fibers at 90$\degree$ angular intervals were placed at each set distance from the center to provide redundant sampling.  The apparatus had a 0.2~mm diameter nylon monofilament suspended vertically in the center of the volume.  The single-mode fiber that carried the laser light was affixed to this central monofilament such that the light was injected into the center of the array.  The first set of WLS fibers were placed 2$\pm$1~mm from the center (the location of light injection), after which each additional set of fibers increased in distance from the center by 4$\pm$1~mm.  Each arm of the fiber array held 8 WLS fibers, at distances of 2, 6, 10, 14, 18, 22, 26, and 30~mm from the center of the volume.  The fiber holder array is shown in Fig.~\ref{fig:fiberHolder}. 

\subsubsection{Calibration}
\label{sec:MAPMT_cal}

\begin{figure}
    \centering
    \includegraphics[width=\textwidth]{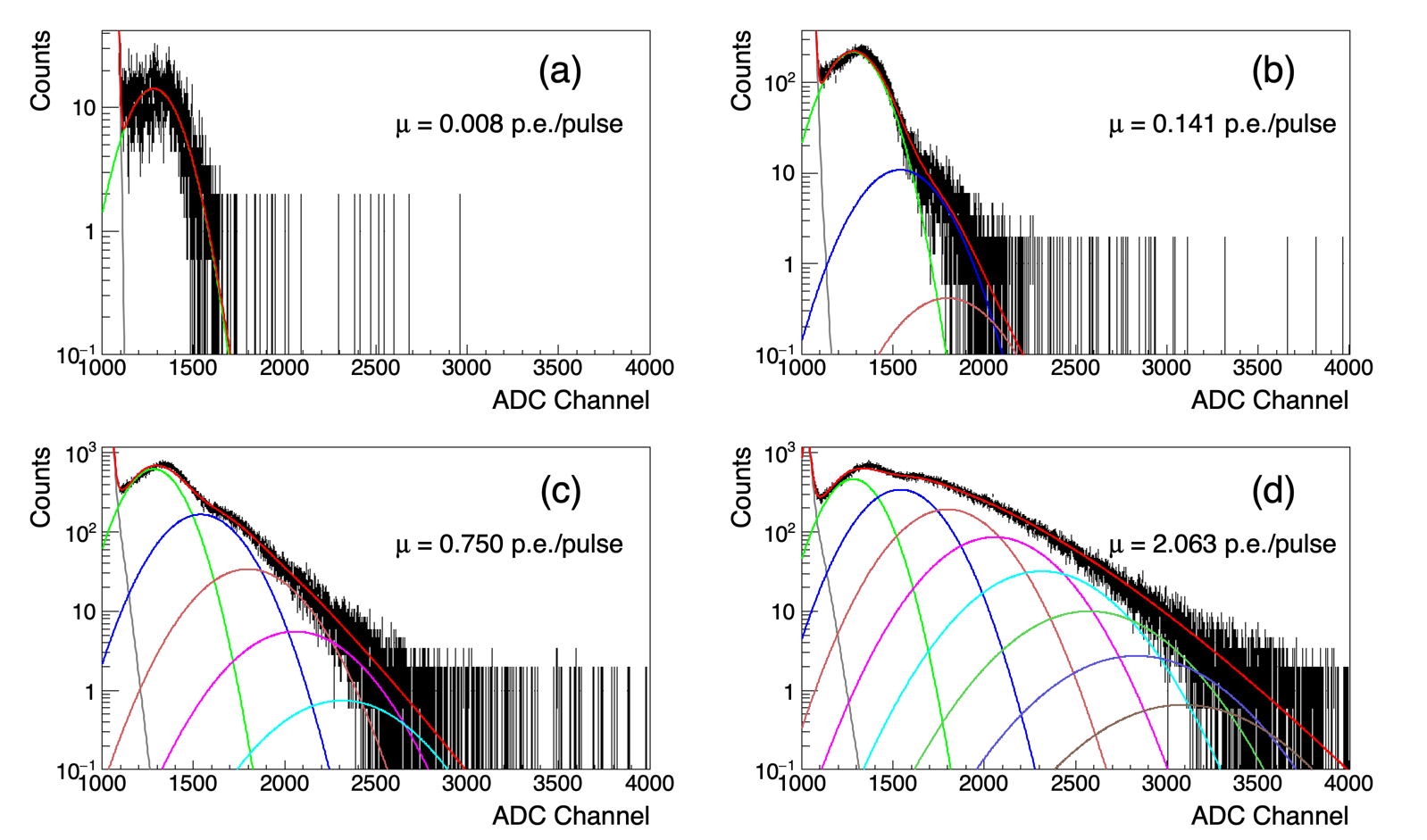}
    \caption{The pulse integral spectra from an example channel of the MAPMT during calibration shown on a semi-log scale.  A constant pedestal charge of 1024 ADC channels was added to aid the fitting process.  (a)~is background, while (b)--(d) show the response to increasing light intensity from a pulsed LED.  In each plot, the red line is the overall fit, while the other colors show the contributions from a single photoelectron (green), two photoelectrons (blue), \textit{etc} (see text for details).  The value $\mu$ shown in each plot is the average photoelectrons per pulse, which was a fitted parameter.}
    \label{fig:PMT_cal}
\end{figure}

\begin{figure}
    \centering
    \includegraphics[width=0.7\textwidth]{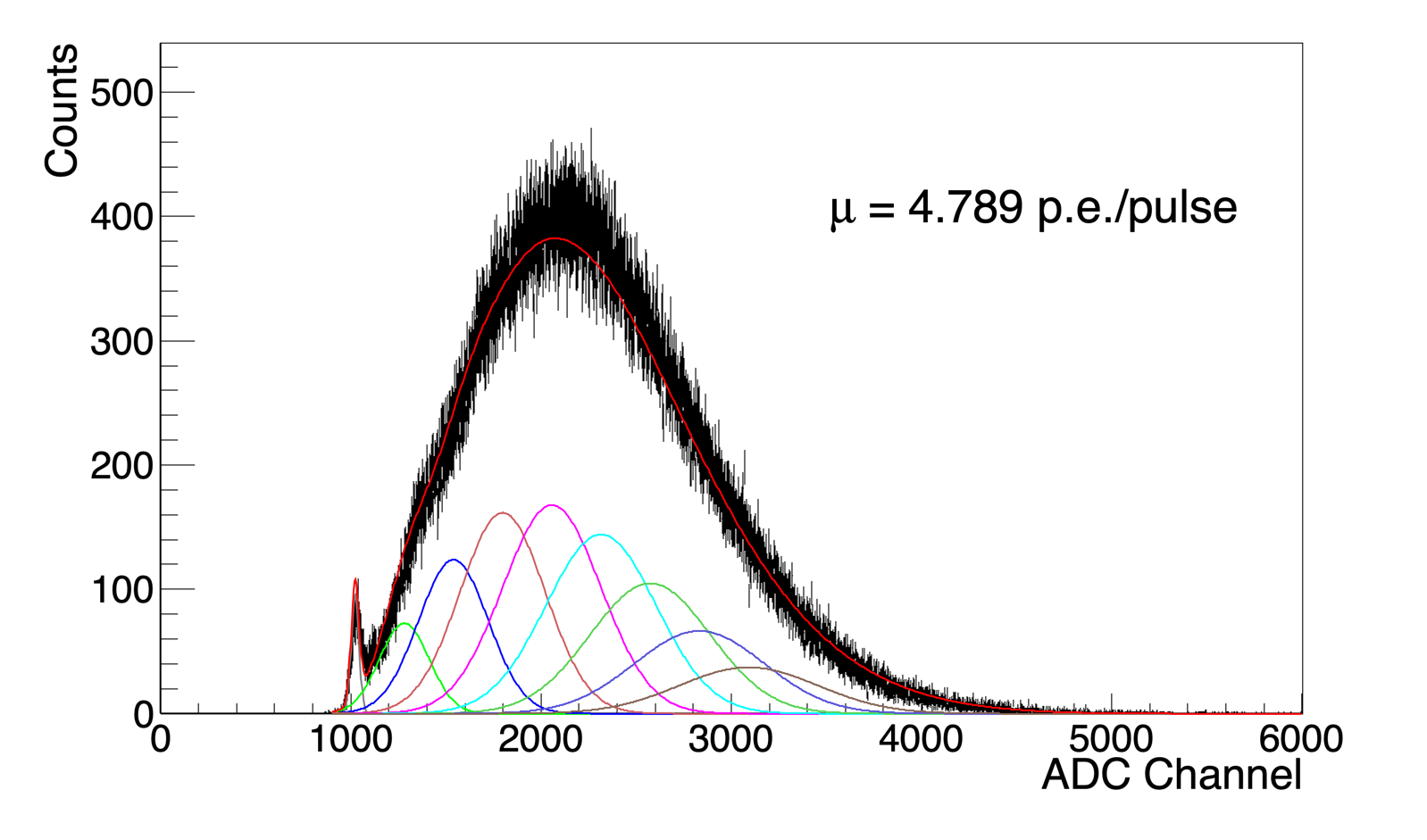}
    \caption{Experimental data from the same channel as in Fig.~\ref{fig:PMT_cal}, fit with the parameters found during calibration.  The red line is the overall fit, while the other colors show the contributions from events in which an increasing number of photoelectrons was produced.  While the events of up to 12 photoelectrons are included in the fitting function, only 8 are shown for clarity. The parameter $\mu$ represents the average number of photoelectrons per pulse.}
    \label{fig:PMT_run}
\end{figure}

In order to determine the number of photoelectrons produced at the photocathode per integration window, we first needed to measure the MAPMT by-pixel response to a single photoelectron.  Prior to each experimental run, we took four calibration measurements with varying intensities of light from a pulsed LED.  The first measurement was a background measurement, in which the signal from a pulser only served to trigger the digitizer.  The pulser signal was also used to drive a 400-nm LED with a pulse width of 3~ns at voltages of 3.5, 3.7, and 3.9~V.  The LED was positioned in a corner of the dark box so it was as far away from the WLS fibers as possible, but within direct view of the portion of the WLS fibers that spann between the opaque liquid vessel and the MAPMT.  In all cases, the light was incident upon the WLS fibers, which were each coupled to their assigned MAPMT pixel.  The resulting four pulse integral spectra were simultaneously fitted according to the method described in Ref.~\cite{Bellamy_1994_PMT_calibration}. The fitting algorithm determines parameters that describe the response of each MAPMT pixel to zero photoelectrons (pedestal charge) and single photoelectrons.  These parameters were then applied to the pulse integral spectra.  Four other parameters, including scaling, two noise parameters, and the average number of photoelectrons per pulse, were allowed to vary.  The fitting algorithm was then applied to the experimental run data, and the average photoelectrons per pulse was recorded.  Example calibration spectra and fits are shown in Fig.~\ref{fig:PMT_cal}, and the final fit to data from an experimental run is shown in Fig.~\ref{fig:PMT_run}.

\subsection{Results}

\begin{figure}
    \centering
    \includegraphics[width=0.6\textwidth]{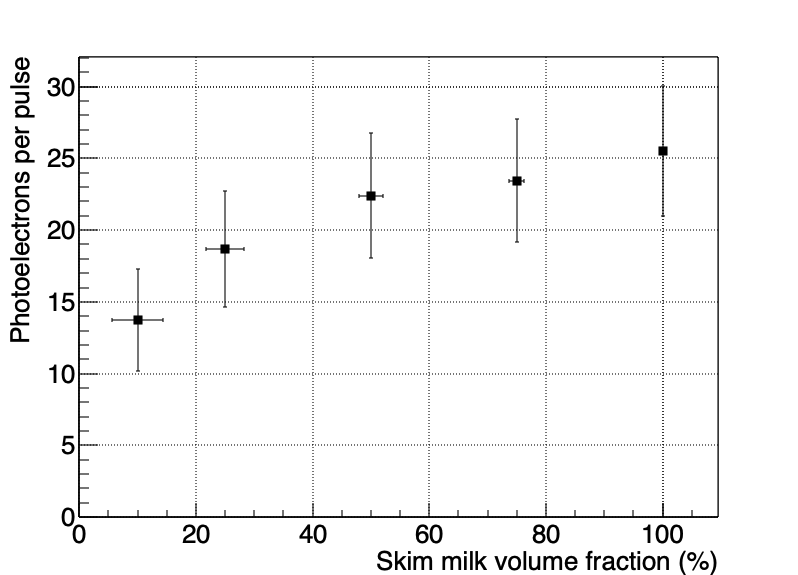}
    \caption{Photoelectrons detected per pulse from photons collected in a WLS fiber closest to the location of light injection.  The first iteration of the experiment was done with 100\% cow skim milk, and then liquid was replaced with distilled water to dilute the milk.  The uncertainty in the reconstructed number of photoelectrons per pulse is derived from the uncertainty of the fitting algorithm described in Sec.~\ref{sec:MAPMT_cal}.  }
    \label{fig:effClose}
\end{figure}

A predicted feature of LiquidO is that light collection can increase with increased scattering for photons that originate close to the WLS fibers.  To demonstrate this experimentally, we injected photons roughly 0.1~mm from the outer cladding of a WLS fiber.  The photons were emitted from a single-mode fiber that was coupled to the pulsed laser described in Sec.~\ref{sec:app}.  The laser was tuned to provide 17,600$\pm$900 photons/pulse.  The single-mode fiber was parallel to the WLS fiber, such that few photons were incident on the WLS fiber without first being scattered.  This was a different arrangement than is shown in Fig.~\ref{fig:fiberHolder}.  Instead of being mounted in the center of the array, the fiber carrying the laser light was affixed directly to the side of a WLS fiber.  The experimental data are presented in Fig.~\ref{fig:effClose} for various dilutions of 0\% fat content (skim) milk, which is used as the scattering medium.  To avoid perturbing the fiber arrangement, the fiber array was not removed from the vessel between iterations of the experiment with different dilutions.  Instead, liquid was siphoned from the volume and replaced with distilled water to increase the dilution.  The new mixture was then stirred for 60~seconds with a stirring rod spun by an electric drill.  As expected, the light collected in the fiber increases with milk concentration. The Spearman correlation coefficient for the data presented in Figure \ref{fig:effClose} is $0.70\pm0.26$, which indicates a strong relationship \cite{Schober2018CorrelationInterpretation}.  As the concentration increases, the scattering length of the medium decreases, and more photons are confined to the region about the WLS fiber and collected.  As the scattering length decreases, the absorption length also decreases; however, light collection efficiency still increases.  Although the average path length of the photons decreases with a higher milk concentration, a greater fraction of that path is in the vicinity of the fiber, which increases the probability of collection.

\begin{figure}
    \centering
    \includegraphics[width=0.6\textwidth]{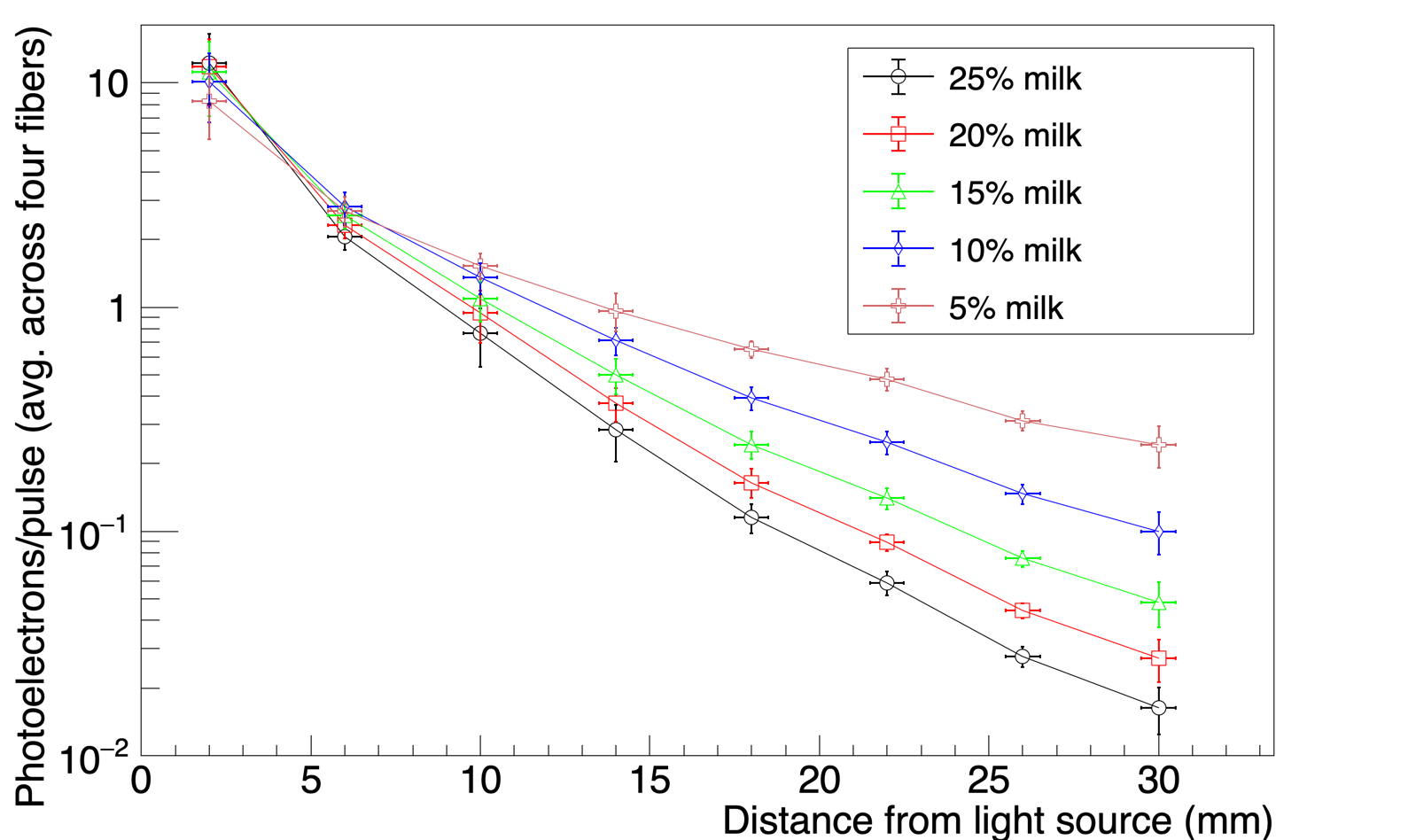}
    \caption{Photoelectrons per pulse from photons collected in WLS fibers at various distances from the light source for a variety of dilutions of commercial, 2\% fat cow milk.  Each pulse of light contained 17,900$\pm$900 photons.}
    \label{fig:lightColl}
\end{figure}

To measure the light collection efficiency into WLS fibers as a function of distance from the light source and bulk liquid opacity, we used the apparatus described in Sec.~\ref{sec:app}.  The opaque liquid used was a dilution of homogenized, commercial, 2\% fat cow milk.  The dilutions given are volumetric mixtures with distilled water (\textit{e.g.}, a 15\% dilution means the mixture was 85\% distilled water by volume).  The laser power for this experiment was 17,900$\pm$900 photons/pulse.  As discussed in Sec.~\ref{sec:app}, there were four fibers at each distance from the center of the fiber array.  The number of photoelectrons per pulse were averaged at each position, and the uncertainty was taken to be the standard deviation of these four measurements.  The results are shown in Fig.~\ref{fig:lightColl}.  Similarly to what was shown in Fig.~\ref{fig:effClose}, the collection efficiency increased with higher concentrations of milk (\textit{i.e.}, more scattering) for the fiber closest to the light source.  Additionally, light collection at distances far from the light source was systematically suppressed by the increased light scattering in higher milk concentrations.  This highlights the localization of optical photons and, therefore, the potential for determining the location of origin of the emitted light.

\section{Modeling}
\subsection{Monte Carlo}
\label{sec:sim}
Our Geant4 model includes the HDPE vessel, the opaque liquid, and the multi-clad WLS fibers in their appropriate locations.  Geant4 has a built-in wavelength-shifting process, which requires user-defined absorption and emission spectra.  The relevant parameters used in the Geant4 model are summarized in Table \ref{tab:ModelParameters}.  The fibers were modeled to have perfectly smooth surfaces, such that reflection and refraction probabilities depended only on solutions to the Fresnel equations given the relative indices of the refraction of the materials in contact.  
\begin{table} [ht]
\centering
\caption{\label{tab:ModelParameters} The material properties used to construct the Geant4 model.}
\begin{tabular}{l|c|c}
    Parameter               & Value                             & Reference \\
    \hline
    Index of refraction (Y-11 outer cladding)   & 1.42 & \cite{KurarayDatasheet}\\
    Index of refraction (Y-11 inner cladding)   & 1.49 & \cite{KurarayDatasheet}\\
    Index of refraction (Y-11 core)             & 1.59 & \cite{KurarayDatasheet}\\
    WLS absorption length                             & See Fig.~\ref{fig:WLSAbsAndEm}  & \cite{Herzkamp2016SimulationNeutrons}\\
    WLS emission spectrum           & See Fig.~\ref{fig:WLSAbsAndEm}  & \cite{Pahlka2019SpectralFibers} \\
    Y-11 bulk absorption length  & See Fig.~\ref{fig:QEandBulkAbs} & \cite{Cooper2007NOAReport} \\
    MAPMT quantum efficiency& See Fig.~\ref{fig:QEandBulkAbs} & \cite{HamamatsuDatasheet} \\
    Reflectivity of HDPE (vessel)   & 0.9                               & \cite{Meinhold2010OpticalColouration} \\
    \hline
\end{tabular}
\end{table}
\begin{figure}
    \centering
    \includegraphics[width=0.7\textwidth]{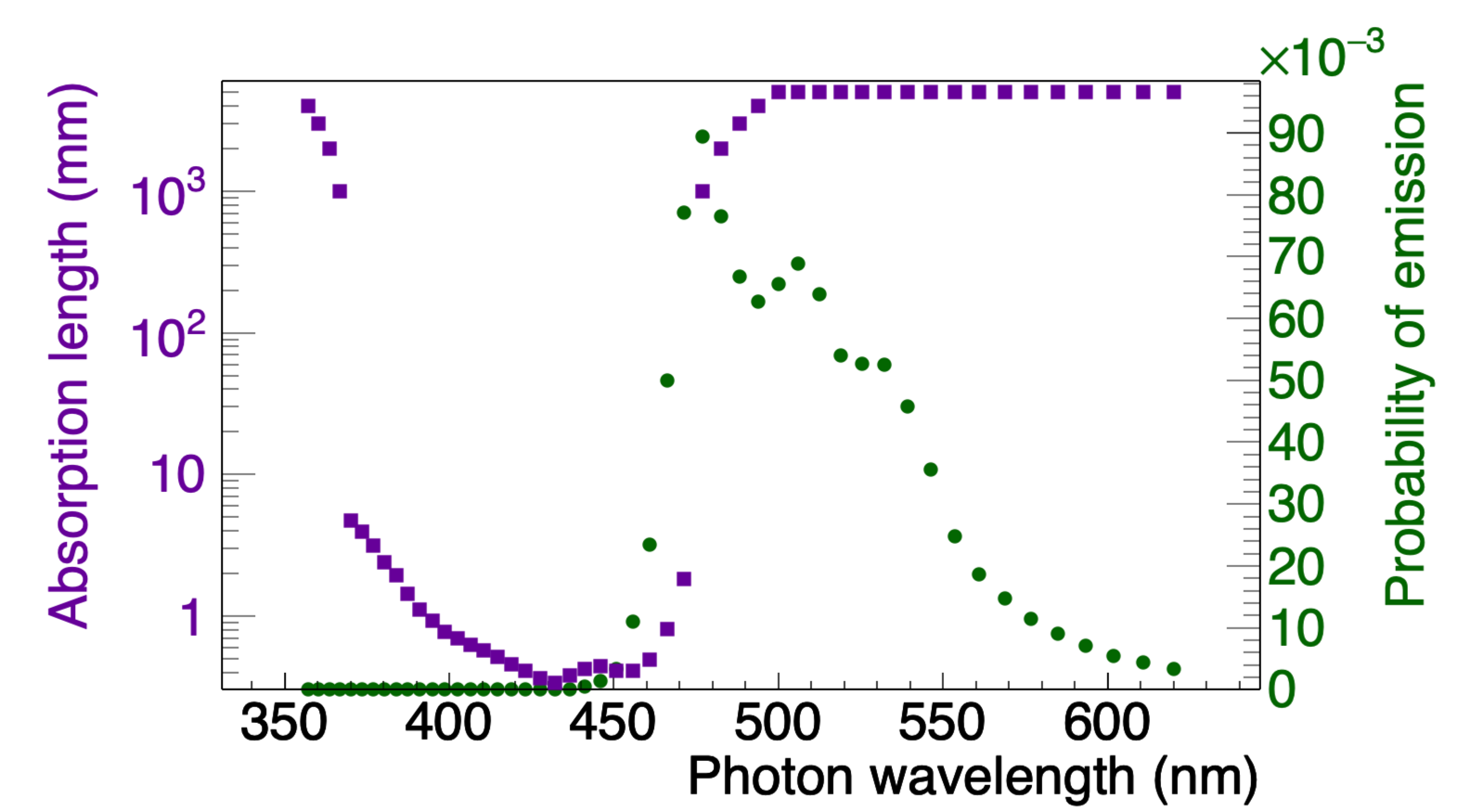}
    \caption{Kuraray Y-11 WLS absorption (purple squares) and emission (green circles) spectra used in the Geant4 model.  WLS absorption results in a new optical photon with a wavelength sampled from the emission distribution and a randomized direction.}
    \label{fig:WLSAbsAndEm}
\end{figure}
\begin{figure}
    \centering
    \includegraphics[width=0.7\textwidth]{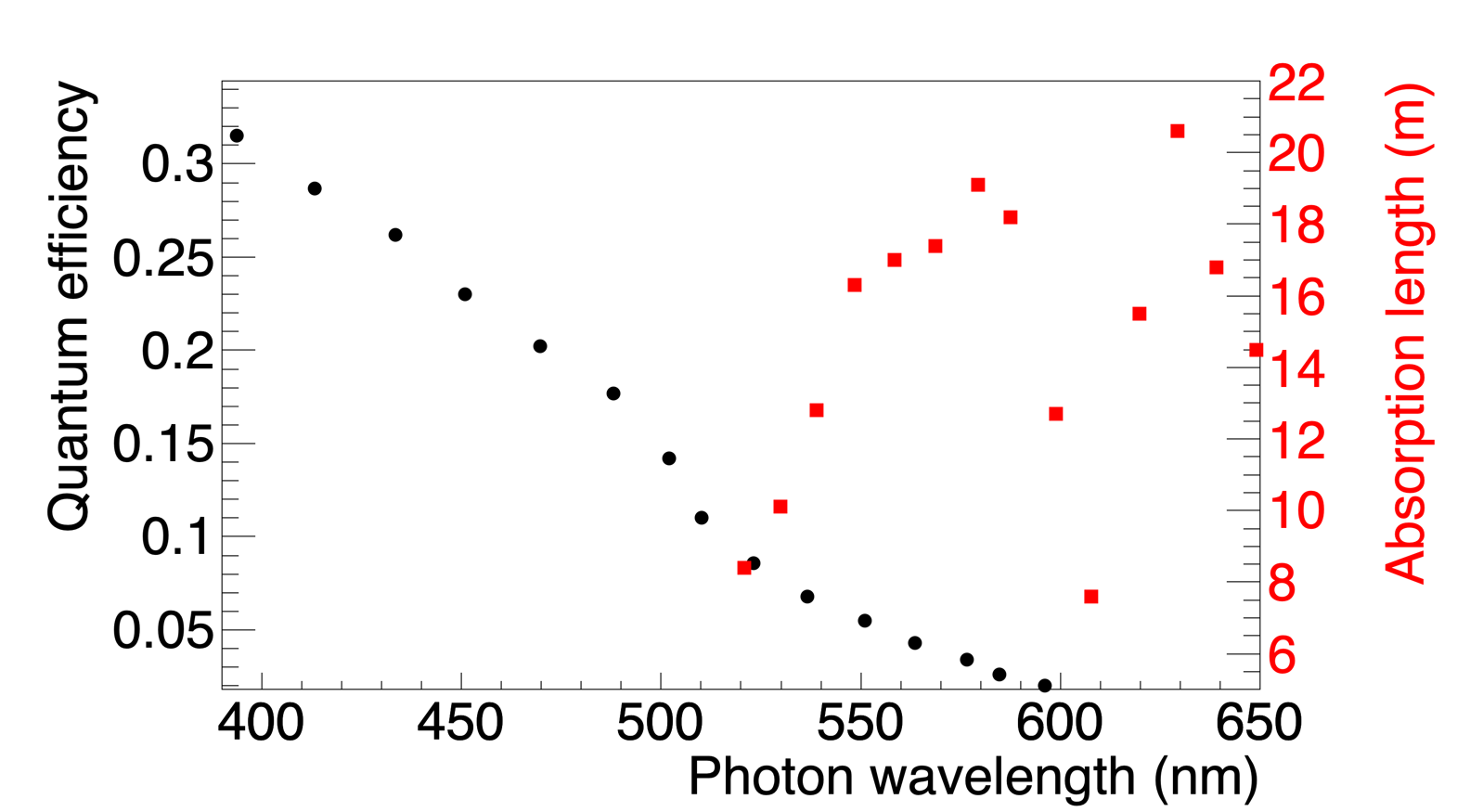}
    \caption{Quantum efficiency (black circles) of the Hamamatsu H12700A and the bulk absorption length of the Kuraray Y-11 fiber (red squares) used in the Geant4 model.  Bulk absorption is a separate process from WLS absorption in Geant4 and cannot result in emission of a new optical photon.  The dip in the absorption length around 610~nm is due to a resonance in the absorption cross-section characteristic to WLS fibers with a polystyrene core~\cite{Cooper2007NOAReport}}
    \label{fig:QEandBulkAbs}
\end{figure}
Cow milk is a complex liquid, with a variable particulate size that depends on fat content, homogenization process, and many other factors.  Typical commercial cow milk has a sufficiently large average particle size for Mie scattering at visible wavelengths~\cite{Stocker2017BroadbandMilk}.  Mie scattering is generally more forward-directed than Rayleigh scattering, which is the approximation for the scattering of photons off of spherical particles that are small in comparison to wavelength.  Another way to account for this anisotropy is to use the reduced scattering coefficient $\mu'_s$:
\begin{linenomath}
\begin{align}
    \mu'_s = \mu_s(1-g).
\end{align}
\end{linenomath}
Here, $\mu_s$ is the scattering coefficient with units of inverse length, and $g$ is the average cosine of the scattering angle; $g=1$ indicates total forward scattering, while $g=0$ represents isotropic scattering. Similarly, the reduced scattering length is
\begin{linenomath}
\begin{align}
    \lambda' = 1/\mu'_s
\end{align}
\end{linenomath}
and has units of length.

The Geant4 framework incorporates the Mie and Rayleigh scattering physics, but we elected to use only Rayleigh scattering because an accurate accounting for Mie scattering requires the knowledge of the particle size distribution, which is unknown for our milk samples.  Geant4 does not allow the specification of an anisotropy parameter, so we used the reduced scattering length in place of the scattering length.  The reduced scattering length and absorption length were treated as free parameters obtained from fit of the model to the experimental data (see Sec.~\ref{sec:compExpSim}.)

Since the number of photoelectrons per pulse is measured in the experiment, the objective is for the model to output results in the same units.  To this end, we included the probability of a photon creating a photoelectron at the photocathode of the MAPMT.  The primary particle in our simulation is a 407-nm optical photon.  In order to result in a photoelectron count for a given channel, the optical photon has to enter a WLS fiber, be refracted through the cladding, absorbed in the core and reemitted in a direction conducive to total internal reflection.  The photon then has to be transported through the 1-m long fiber, during which it could be absorbed by the fiber core.  At the end of the fiber, the photon undergoes conversion to photoelectron at the photocathode with a quantum efficiency that depends upon its wavelength.  The values for the wavelength-dependent quantum efficiency were taken from the manufacturers datasheet and are shown in Fig.~\ref{fig:QEandBulkAbs}. 

\label{sec:compExpSim}

The optical parameters of an actual candidate LiquidO media should be well-characterized prior to prototyping.  However, as the determination of the optical parameters of our milk samples was not the purpose of our study, we did not directly measure them.  Instead, to find the best-fit parameters for the reduced scattering length and absorption length of our milk sample, we made measurements at a series of dilutions.  The absorption coefficient $\mu_{a,c}$ of a compound can be calculated from
\begin{linenomath}
\begin{align}
    \frac{\mu_{a,c}}{\rho_c} = \sum_i w_i \frac{\mu_{a,i}}{\rho_i}
\end{align}
\end{linenomath}
where $\rho$ is the density of the material, and $w_i$ is the fraction of the $i$th material by mass.  The reduced scattering length can be calculated in an analogous way under the assumption that the particle size distribution is unaffected by concentration.  The validity of this assumption is discussed below.

\begin{figure}
    \centering
    \includegraphics[width=0.5\textwidth]{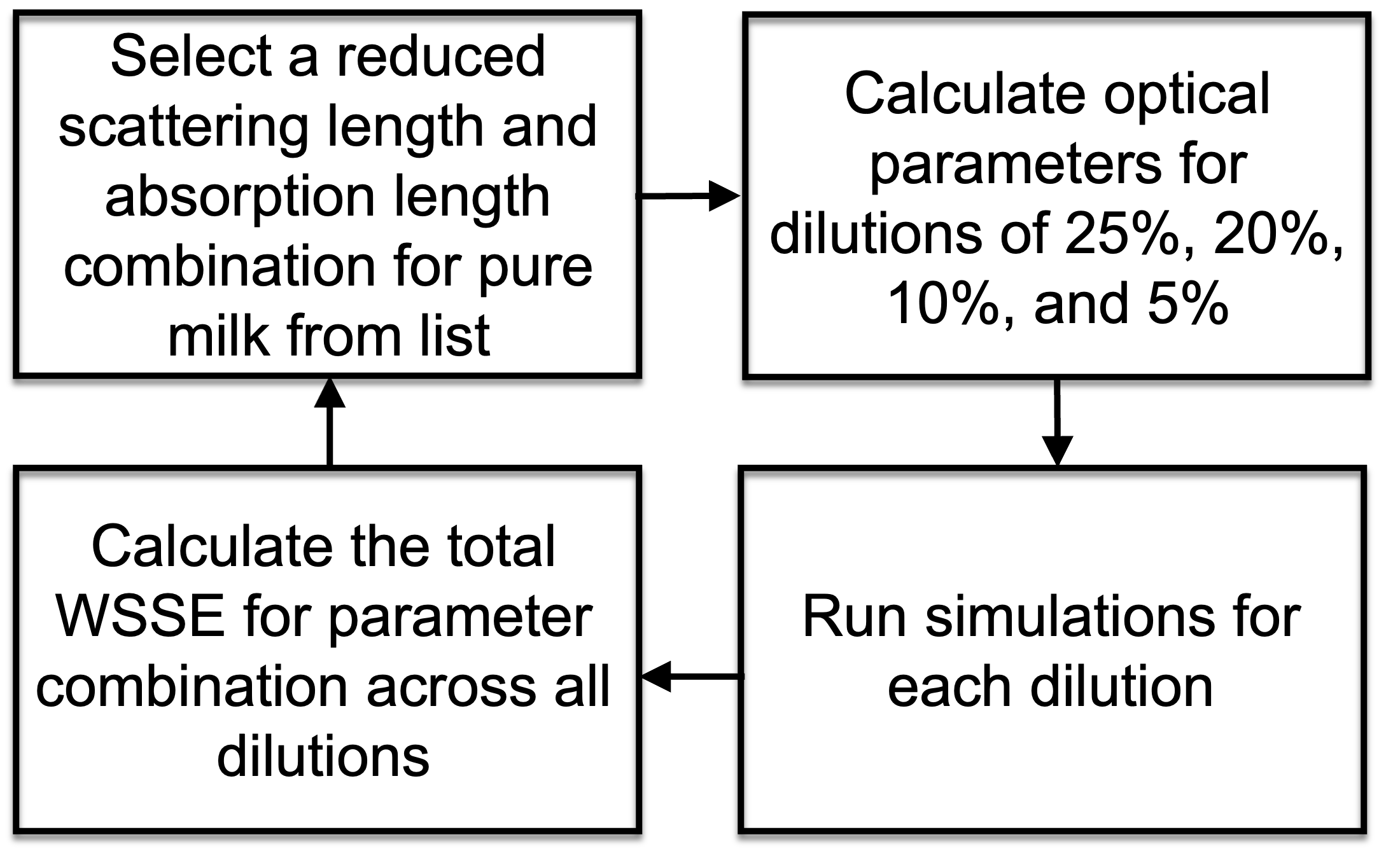}
    \caption{A flowchart of the parameter search process to determine the optical parameters of pure milk.}
    \label{fig:flowChart}
\end{figure}

In the parameter search, we chose a combination of reduced scattering and absorption lengths for pure milk, calculating the resulting parameters and ran a simulation for each dilution; we then calculated the total weighted sum of squared errors (WSEE):
\begin{linenomath}
\begin{align}
    \textnormal{WSSE} = \sum_{i=1}^8 \frac{1}{\sigma_i^2}(y_{i,\textnormal{sim}}-y_{i,\textnormal{exp}})^2,
\end{align}
\end{linenomath}
where $\sigma_i$ is the uncertainty of the measurement at the $i$th distance from the light source, and $y_{i,\textnormal{sim}}$ and $y_{i,\textnormal{exp}}$ are the average numbers of photoelectrons from the four fibers at that distance from the light source in the simulation and experiment, respectively. The values of parameters were pre-selected at regular intervals such that the search comprised a grid.  This process is shown as a flowchart in Fig.~\ref{fig:flowChart}.

\begin{figure}
    \centering
    \includegraphics[width=\textwidth]{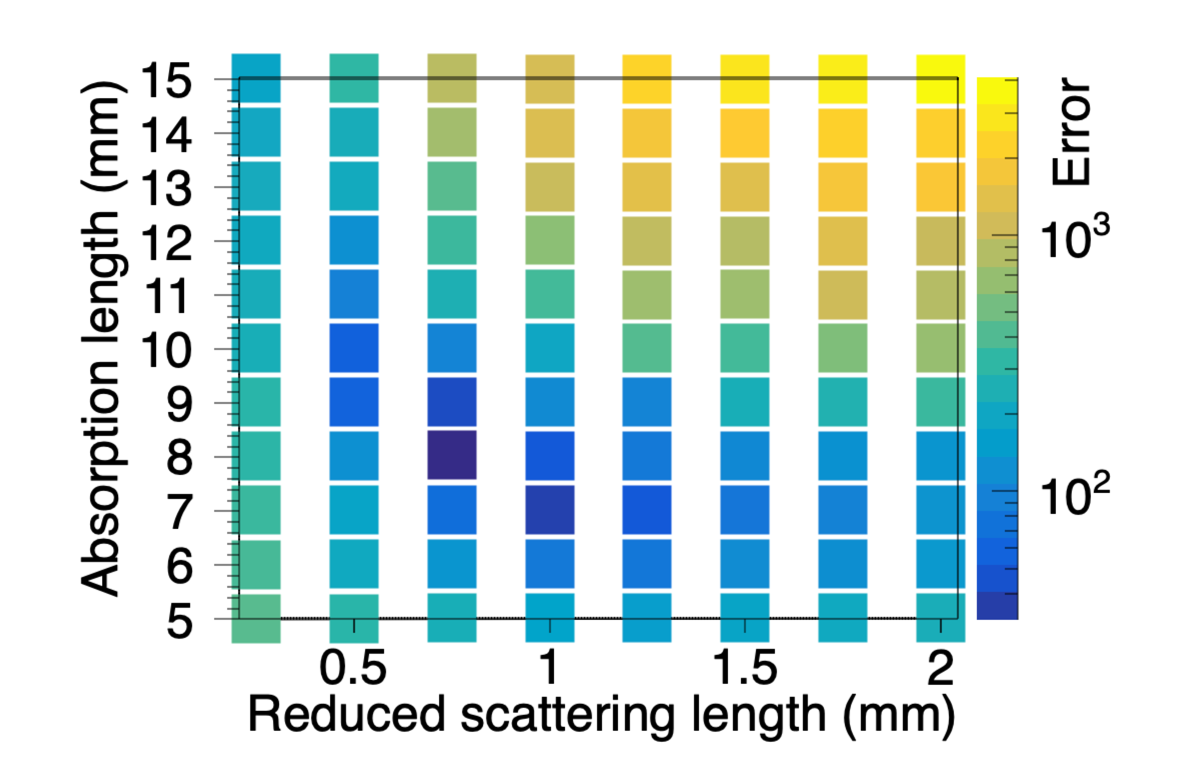}
    \caption{The total WSSE values for combinations of scattering and absorption parameters for simulation and experimental data.  The total WSSE is the individual WSSEs from the 25\%, 20\%, 10\%, and 5\% dilutions of 2\% milkfat commercial milk datasets, added in quadrature.  The minimum error was 31.7, at a reduced scattering length of 0.75~mm and an absorption length of 8~mm.}
    \label{fig:WSSETotal}
\end{figure}

The dataset from 15\% dilution was withheld from the fitting for validation. The results of the parameter search are shown in Fig.~\ref{fig:WSSETotal}.  The parameter combination with the lowest total error (31.7) is a reduced scattering length of 0.75~mm and an absorption length of 8~mm.  

\begin{figure}
    \centering
    \includegraphics[width=\textwidth]{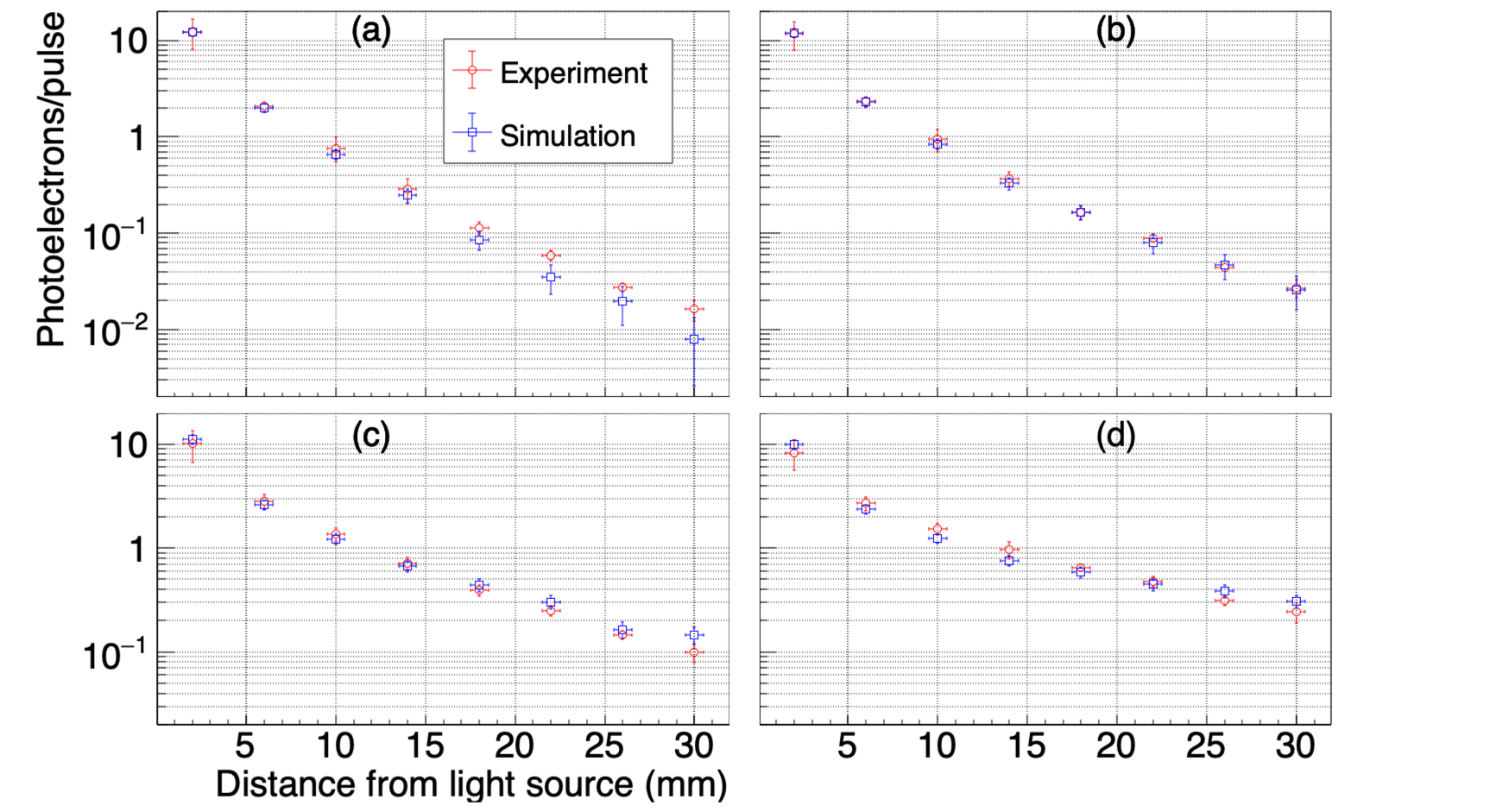}
    \caption{Comparison of simulated and experimental results for (a)~25\% milk, (b)~20\% milk, (c)~10\% milk, and (d)~5\% milk.  All simulated values agree with experiment within uncertainty.  The optical parameters used for pure milk were optimized as shown in Fig.~\ref{fig:WSSETotal}.  Note that the vertical scale for (a) and (b) differs from the scale for (c) and (d).}
    \label{fig:expSimCompAll}
\end{figure}

The comparison between experimental and simulated data for these parameters is shown in Fig.~\ref{fig:expSimCompAll}.  The $\chi^2$ values are 26.3, 2.4, 11.3, and 13.6 for concentrations of 25\%, 20\%, 10\%, and 5\%.  The fits are best for concentrations of 20\% and 10\%, and worst for concentrations of 25\% and 5\%.  The particle size distributions for dilutions of milk above 10\% are inconstant, which introduces nonlinearity into the rate at which the reduced scattering length changes with concentration \cite{Urban1999ApplicationMilk}.  One study found a measureable departure from reduced scattering length linearity in milk concentrations above approximately 0.6\% \cite{Campbell2007ScalingScatterers}.  The best overall fit found by our parameter search approximates an average for the particle size distribution, which likely explains why the fits degrade in quality away from the average concentration. 

\begin{figure}
    \centering
    \includegraphics[width=0.7\textwidth]{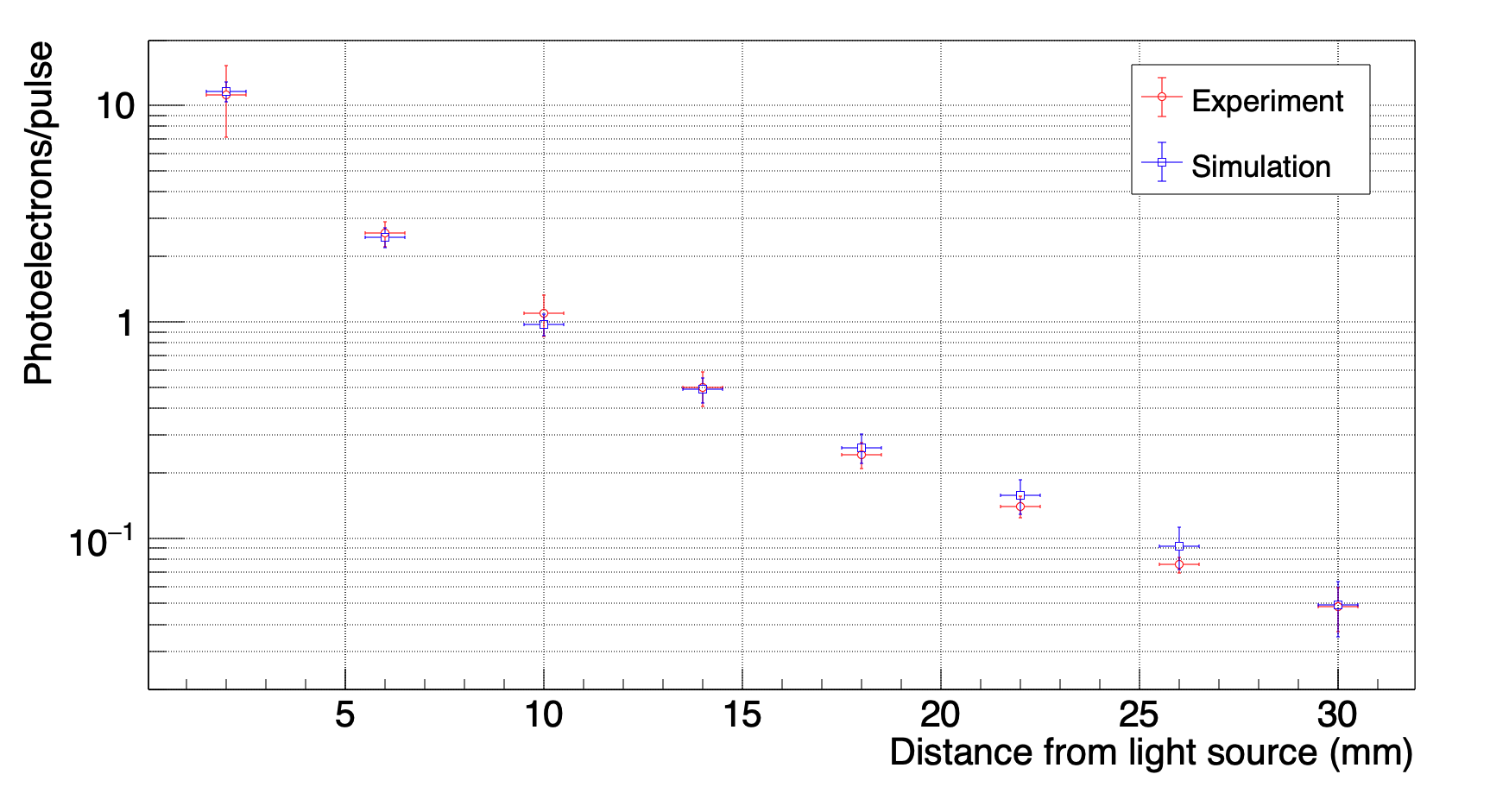}
    \caption{The comparison of simulated and experimental results for 15\% milk dilution.  The optical parameters for milk used were the same as shown in Fig.~\ref{fig:expSimCompAll}.}
    \label{fig:expSimComp15}
\end{figure}

As a validation of the selected optical parameters, we compared the dataset for 15\% milk dilution (which was omitted from the parameter search) to simulation; the results are in Fig.~\ref{fig:expSimComp15}.  The data shows good agreement ($\chi^2 = 9.0$) with simulation at all distances from the origin.  Summing the $\chi^2$ values from all five datasets gives a $\chi^2$ value of 62.5, which corresponds to a reduced $\chi^2$ of 1.60, and a p-value of 0.013.

Ref.~\cite{Qin2005HyperspectralJuices} measured the absorption and reduced scattering lengths for 2.44\% fat cow milk in the wavelength range of 530--900~nm.  Extrapolating their data to 407~nm gives an absorption length of approximately 7~mm  and a reduced scattering length of 0.8~mm, very close to the values obtained from the fit to experimental data.  Ref.~\cite{Stocker2017BroadbandMilk} found that the reduced scattering length of whole cow milk at 400~nm can vary in the range of 0.3--1.0~mm depending on homogenization time, and the absorption length at 400~nm is as high as 40~mm for 1.5\% milkfat.  

\subsection{Diffusion Approximation}

Liquid-O type detectors are characterized by a bulk medium with a short scattering length and long attenuation length.  Within this medium there is an array of wavelength shifting fibers acting as photon sinks.  These detector requirements provide conditions for diffusion calculations to approximately describe photon dynamics in the system and provide verification of the Monte Carlo model.  


The diffusion equation can be derived by simplifying the Boltzmann transport equation; this is common in modeling neutron transport~\cite{Duderstadt1976NuclearAnalysis}.  We assume photon flux is linearly anisotropic, photon current varies slowly compared to mean collision time, and our sources are isotropic.  This can be physically realized in a Liquid-O type detector by spacing fibers and sources many scattering lengths apart on a uniform grid.

Here, we aim to predict the probability of photon capture in the WLS fiber given an initial photon distribution.
We are most concerned with three processes the photon may undergo: scattering in the bulk medium, absorption in the bulk medium, and absorption in the fibers.  We use two equations to describe two groups of photons; the first tracks a photon density in the bulk medium, while the second tracks the absorbed photon density in the wavelength shifting material:
\begin{linenomath}
\begin{align}
    \frac{\partial \rho_m}{\partial t} = &
    \nabla \left[ D \nabla \phi_m \right] - \frac{\phi_m}{\alpha_m} -\frac{\phi_m}{\alpha_f}, \\
    \frac{\partial \rho_f}{\partial t} = &
    \frac{\phi_m}{\alpha_f}, \\
    D = &
    \frac{1}{3(\alpha_m^{-1}+\alpha_f^{-1}+\sigma_m^{-1})}.
\end{align}
\end{linenomath}
Here, $\rho_m$ and $\rho_f$ are photon densities in the bulk material and WLS fiber, respectively, and $\phi_m$ is the photon flux in the medium.  The material properties pertinent to the model are $\alpha_m$, the mean free path to absorption in bulk material, $\alpha_f$, the mean free path to absorption in WLS fiber, and $\sigma_m$, the mean free path to scattering in bulk material. Finally, $t$ is time, and $D$ is the diffusion coefficient.

Using the speed of light in the medium, we can replace $\phi = \rho c/n$ and rewrite the coupled diffusion equations entirely in terms of photon density.  Assuming $n_{\text{fiber}}\approx n_{\text{medium}}$, we have:
\begin{linenomath}
\begin{align}\label{eq:diffeqn}
    \frac{\partial \rho_m}{\partial t} = &
    \frac{c}{n} \left( \nabla \left[ D \nabla \rho_m \right] - \frac{\rho_m}{\alpha_m} -\frac{\rho_m}{\alpha_f}\right) \\
    \frac{\partial \rho_f}{\partial t} = &
    \frac{c}{n}\frac{\rho_m}{\alpha_f};
\end{align}
\end{linenomath}
where $c$ is the speed of light in vacuum, and $n$ is the index of refraction in the medium.


To verify the Monte Carlo simulation, we generate a uniform square grid of WLS fibers with a pitch of 15~mm, a scattering length of 1~mm and an absorption length of 50~mm for the bulk medium.  An isotropic source was placed at the center of this square array of fibers.  We solve Eq.~(\ref{eq:diffeqn}) for this geometry using py-pde, a library implementing the finite difference method~\cite{Zwicker2020Py-pde:Equations}.  The mesh size of the simulation was set to 0.05~mm, an order of magnitude smaller than the absorption length of the fibers.  Von Neumann stability analysis \cite{Charney1950NumericalEquation} is used to determine an upper bound on $\Delta t$ for numerical stability of the solution:
\begin{linenomath}
\begin{equation}
    \Delta t \leq \frac{n}{2Dc} \left( \frac{1}{(\Delta x)^2} + \frac{1}{(\Delta y)^2} \right)^{-1} \approx 6 \textnormal{ fs},
\end{equation}
\end{linenomath}
where $\Delta x$ and $\Delta y$ are the mesh sizes in the x and y directions.
To increase computational efficiency, we leverage two types of symmetry in the detector model. First, the detector is homogeneous in the axial direction, and the medium boundaries are many absorption lengths away from the source, allowing us to treat the detector as infinite. Since the photon distribution along the axis of the fiber is of less interest, we can incorporate this symmetry by solving the diffusion equation in two dimensions.  Second, the detector and source contain symmetries of the dihedral group $D_4$.  Of the eight symmetries available to us, we take advantage of four by modeling one quadrant in the $x$\nobreakdash-$y$~plane with Neumann boundaries at the quadrant boundaries. The last set of symmetries could be implemented by slicing each quadrant diagonally through the origin, but we did not take advantage of this symmetry due to the increased complexity of implementation.

\begin{figure}
    \centering
    \includegraphics[width=0.7\textwidth]{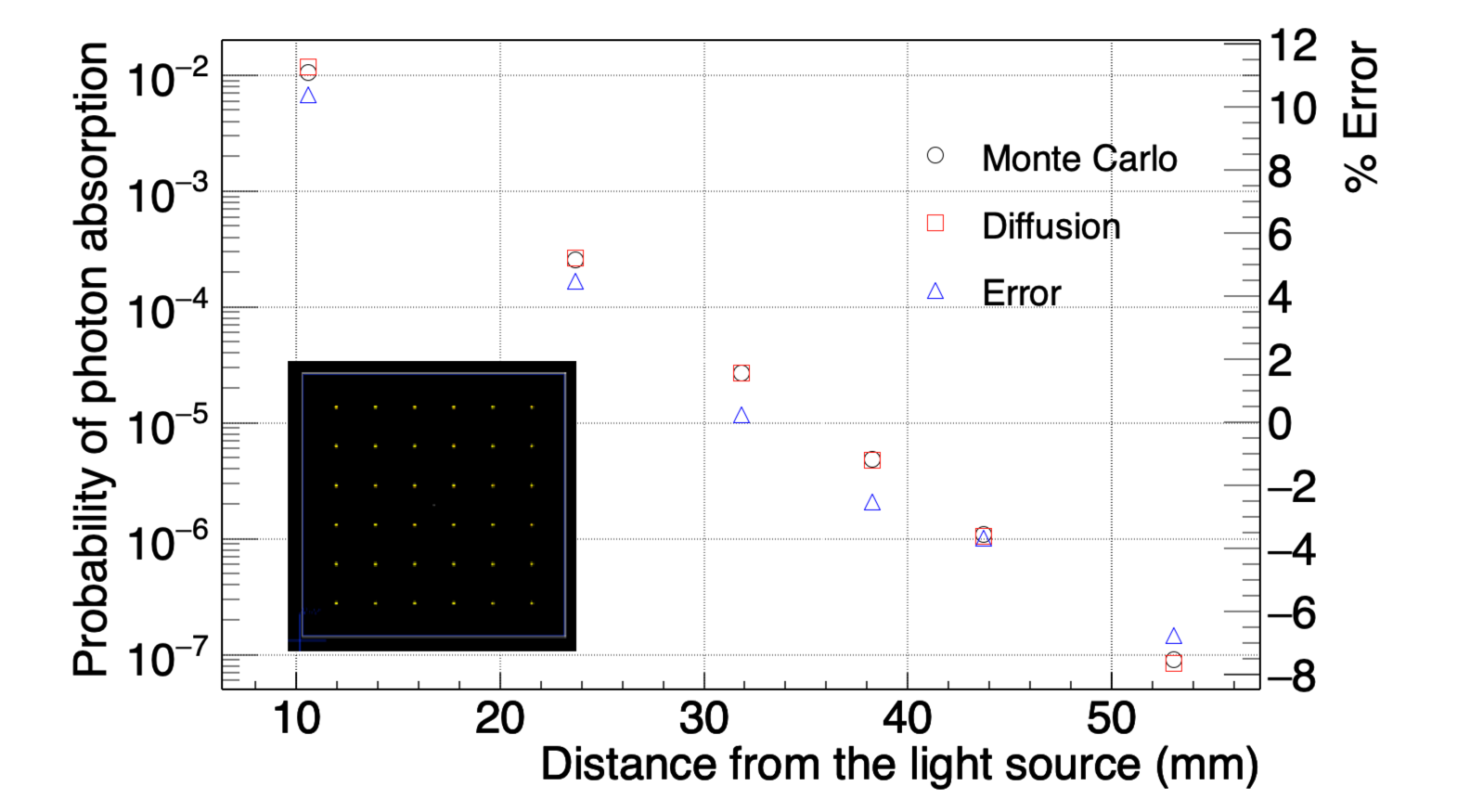}
    \caption{Photon absorption probabilities for a square grid array of WLS fibers with a fiber pitch of 15~mm simulated using the diffusion approximation and Monte Carlo.  The inset shows a visualization of the geometry in Geant4.  The two simulation methods perform within 15\% of each other at each fiber.}
    \label{fig:diff_compare}
\end{figure}

For the square grid geometry, we calculate the probability of each WLS fiber absorbing a photon using both the diffusion and Monte Carlo models.  The results for each are shown in Fig.~\ref{fig:diff_compare}; they agree within 11\% for the diffusion and Monte Carlo simulation for each fiber.

\section{Extension of model to low-absorption medium}

The results described in this study pertain to dilutions of cow milk, which was chosen because it is readily available, water-soluble, and highly scattering.  However, the relatively short absorption length of $<$10~mm means that the overall efficiency of photon collection is poor.  In contrast, NoWASH can be created with an absorption length of above 2~m, while still maintaining a scattering length of below 1~mm~\cite{Buck2019}.  WbLS has an even longer absorption length of tens of meters~\cite{Bignell2015CharacterizationScintillator}, but measurements of the scattering length for opaque WbLS have not yet been carried out.  

\begin{figure}
    \centering
    \includegraphics[width=0.7\textwidth]{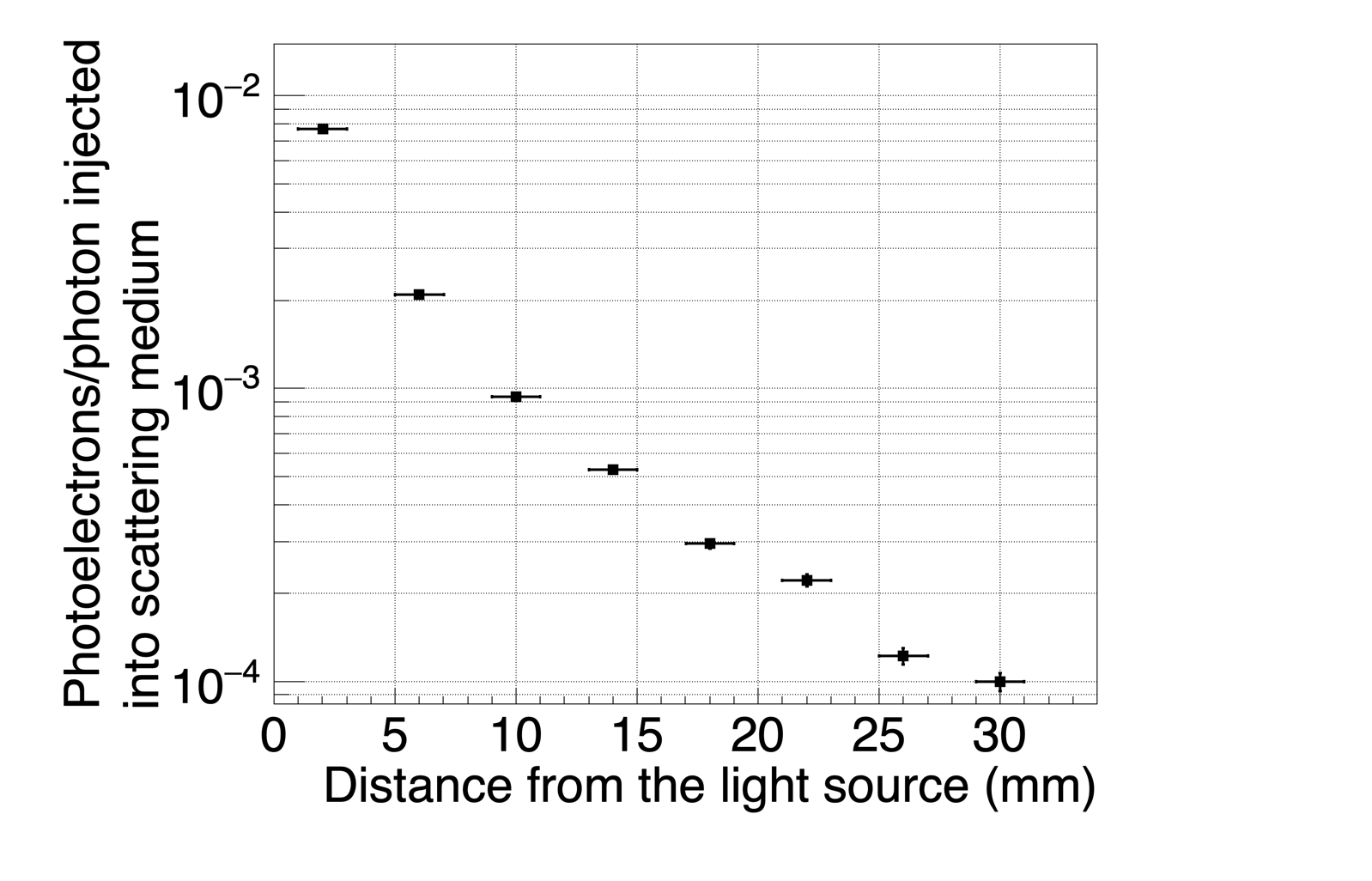}
    \caption{Simulated efficiency of light collection possible in an opaque liquid with a 5~m absorption length and a 0.5~mm reduced scattering length.  The plot shows the number of photoelectrons detected per photon injected into the system, and includes losses in transport in the bulk medium, collection in the WLS fiber, and conversion at the photocathode.}
    \label{fig:highEff}
\end{figure}

As an illustration of the efficiencies possible in a real detector with desirable optical properties, we repeated our simulation from Sec.~\ref{sec:sim} with a reduced scattering length of 0.5~mm and an absorption length of 5~m, with a single WLS fiber at each distance from the origin.  We also modeled light collection at both ends of the WLS fibers instead of only one end as in the experiment; the results are shown in Fig.~\ref{fig:highEff}.  In this instance, the fiber 2~mm from the light source produces approximately 0.0077 photoelectrons for each incident photon.  WbLS with a 10\% loading of LAB-PPO has a light yield of 1357$\pm$125 photons/MeV~\cite{Caravaca2020CharacterizationSeparation}; therefore, an event that deposits 1~MeV of energy 2~mm from a WLS fiber in 10\% loaded WbLS with the optical parameters discussed above could be expected to produce about 10 photoelectrons, which is readily detectable in this experiment. The efficiency would be further increased by instituting a lattice of WLS fibers, such that light was collected from a single event in multiple fibers. Additionally, the MAPMT could be substituted for a silicon photomultiplier (SiPM) which can have a higher quantum efficiency, though at a cost of increased dark noise at room temperature.  We note that this is an extension of our model to conditions that were not directly experimentally validated.  While the basic physics is unchanged for shorter scattering lengths and longer absorption lengths, these conditions will need to be confirmed as part of future detector validation.

\section{Conclusion}
\label{sec:conclusion}

In this work, we quantified the light coupling efficiency from an opaque liquid (dilutions of commercial cow milk) into a WLS fiber as a function of the optical parameters of the liquid and the distance from the origin of the light to the fiber.  We validated a Geant4 model against our experimental data and demonstrated agreement between the Geant4 model and a simplified diffusion approximation to within 11\% error.  We used the validated model to show the efficiencies that may be possible in LiquidO-style detectors of optimal optical properties.
The next step is to use this validated model to design a detector using a scintillating liquid to detect ionizing radiation.  We intend to use our model to optimize a detector design for the reconstruction of event topology.  We hope to use event topology to expand the applications of this technology to source localization, particle-type identification, and high levels of doping (to increase neutron detection efficiency). 

\section*{Acknowledgements}
This work was supported in part by Department of Energy National Nuclear Security Administration, Consortium for Monitoring, Technology, and Verification (DE-NA0003920); Department of Defense, Defense Threat Reduction Agency (HDTRA1-20-2-0002); and high-performance computer time and resources from the DoD High Performance Computing Modernization Program.

\bibliographystyle{model1-num-names}
 \bibliography{references}

\end{document}